\documentclass[twocolumn,prd,nofootinbib]{revtex4}

\newif\ifpdf
\ifx\pdfoutput\undefined
\pdffalse % we are not running PDFLaTeX
\else
\pdfoutput=1 % we are running PDFLaTeX
\pdftrue
\fi

\ifpdf
\usepackage[pdftex]{graphicx}
\else
\usepackage{graphicx}
\fi

\usepackage{epsf}% Include figure files
\usepackage{dcolumn}% Align table columns on decimal point

\newcommand{\be}{\begin{equation}}
\newcommand{\ee}{\end{equation}}

\def\4he{$^4$He}
\def\3he{$^3$He}
\def\7li{$^7$Li}

\newcommand\la{\lower0.6ex\vbox{\hbox{\ensuremath{\buildrel{\textstyle<}\over{\sim}\ }}}}
\newcommand\ga{\lower0.6ex\vbox{\hbox{\ensuremath{\buildrel{\textstyle>}\over{\sim}\ }}}}

\def\lsim{\mathrel{\raise.3ex\hbox{$<$\kern-.75em\lower1ex\hbox{$\sim$}}}}
\def\gsim{\mathrel{\raise.3ex\hbox{$>$\kern-.75em\lower1ex\hbox{$\sim$}}}}

\begin{document}
 \ifpdf
\DeclareGraphicsExtensions{.pdf,.jpg,.mps,.png}
 \else
\DeclareGraphicsExtensions{.eps,.ps}
 \fi

\title{LSND anomaly from $CPT$ violation
in four-neutrino models}
\author{V. Barger$^{1}$, D. Marfatia$^{2}$ and K. Whisnant$^{3}$}
\affiliation{$^1$Department of Physics, University of Wisconsin, Madison, WI 53706}
\affiliation{$^2$Department of Physics, Boston University, Boston, MA 02215}
\affiliation{$^3$Department of Physics, Iowa State University, Ames, IA 50011}

\begin{abstract}

The LSND signal for $\bar\nu_\mu \to \bar\nu_e$ oscillations has
prompted supposition that there may be a fourth light neutrino or that
$CPT$ is violated. Neither
explanation provides a good fit to all existing neutrino data. We
examine the even more speculative 
possibility that a {\it four-neutrino model with $CPT$
violation} can explain the LSND effect and remain consistent with all
other data. We find that models with a 3~+~1 mass structure in the
neutrino sector are viable; a  2~+~2 structure
is permitted only in the antineutrino sector.

\end{abstract}

\maketitle

\section{Introduction}

The LSND experiment has found evidence for $\bar\nu_\mu \to \bar\nu_e$
oscillations at the $3.3\sigma$ level~\cite{lsnd-dar, lsnd-final}, with
indications for $\nu_\mu \to \nu_e$ oscillations at lesser
significance~\cite{lsnd-final, lsnd-dif}. The combination of the LSND
data with the compelling evidence for oscillations in
solar, atmospheric, accelerator, and reactor neutrino
experiments
cannot be adequately explained in the standard
three-neutrino picture with $CPT$ 
conservation~\cite{sterile}.
Extensions to models with four light neutrinos (with the extra
neutrino being sterile)~\cite{sterile} or $CPT$ violation~\cite{cptv,
cptv2} with three neutrinos have been proposed to accommodate all
neutrino data. However, in both cases, recent analyses
indicate that neither scenario provides a good 
description of the data~\cite{not-four, not-cptv}.

The MiniBooNE experiment~\cite{miniboone} is now taking data that will
test the LSND oscillation 
parameters\footnote{The bulk of the parameter region 
allowed by $\bar\nu_\mu \to \bar\nu_e$ data from
LSND and KARMEN~\cite{karmen} and  $\bar\nu_e \to \bar\nu_e$ data
from the Bugey reactor experiment~\cite{bugey}, lies in a
narrow band in $(\sin^22\theta_L, \delta m^2_L)$ space along the
line described approximately by $\sin^22\theta_L (\delta m^2_L)^{1.64}
= 0.0025$ between $\delta m^2_L \sim 0.2$ and 1~eV$^2$. A small
allowed region near $\delta m^2_L \sim 7$~eV$^2$ and 
$\sin^22\theta_L = 0.004$ also exists~\cite{church}.} 
in the $\nu_\mu \to
\nu_e$ channel. A positive result in MiniBooNE will rule out current
versions of $CPT$-violating models, while a negative result will
rule out four-neutrino models with $CPT$ conservation. In either case,
the surviving models will still not give a good fit to all data.

In this letter we consider the very speculative possibility of {\it
$CPT$ violation in four-neutrino models}. The $CPT$ violation is
manifested as different mass matrices for neutrinos and
antineutrinos\footnote{Whether such a model can be constructed 
using nonlocality
of the interactions without violating Lorentz invariance 
is still a matter of debate~\cite{greenberg}.}. 
We find that such a scenario can accommodate the data only if the 
sterile neutrino is weakly coupled to
active neutrinos. 
Thus, while 3~+~1 models are viable, 2~+~2 models (in which the sterile
neutrino is strongly coupled to active neutrinos in solar and/or
atmospheric oscillations), are not. 
A hybrid solution (3~+~1 for neutrinos and
2~+~2 for antineutrinos) is also possible.

\section{Four neutrinos or $CPT$ violation?}

There are two types of four-neutrino models: (a) 3~+~1, where 
active neutrinos
have mass-squared differences and mixings similar to the standard
three-neutrino model that describes solar and atmospheric data, and
(b) 2~+~2, where there are two pairs of closely spaced mass
eigenstates, one of which accounts for the solar neutrino data and the
other for the atmospheric neutrino data. The 3~+~1
models with $CPT$ conservation are disfavored because the Bugey
reactor~\cite{bugey} and CDHSW accelerator~\cite{cdhsw} experiments
put constraints on oscillation amplitudes for $\bar\nu_e \to
\bar\nu_e$ and $\nu_\mu \to \nu_\mu$ survival, respectively, which
together imply an upper limit on the LSND oscillation amplitude that
is below the experimental value~\cite{not-four, 4nu-constraints,
our-four}. The 2~+~2 models with $CPT$ conservation are ruled out
because the combination of solar and atmospheric data do not allow
enough room for a full sterile neutrino~\cite{not-four}. 
To resurrect the scenario in which sterile and active states 
are only weakly coupled, it has been shown that extending 3~+~1
models by an extra sterile neutrino improves the fit
to short-baseline data substantially~\cite{sorel}. 
The constraints on 2~+~2 models may be relaxed by 
including certain small neglected mixing angles in the 
analysis~\cite{pas-song-weiler}.  

In $CPT$-violating models with
three neutrinos, the mass spectra, and hence the mass-squared
differences, are different for neutrinos and 
antineutrinos\footnote{A comparison of solar and reactor
neutrino data can constrain $CPT$ violation~\cite{us}; a 90\% C.~L. limit of
$|\delta \bar m^2 -\delta m^2| < 1.3\times 10^{-3}$ eV$^2$
was found in Ref.~\cite{murayama}.}. In the
original versions of these models~\cite{cptv}, 
the neutrino mass-squared differences
accounted for the solar and atmospheric oscillations (but not the weak
$\nu_\mu \to \nu_e$ signal in LSND), while the antineutrino
mass-squared differences accounted for the LSND and atmospheric
oscillations. With the addition of the KamLAND data indicating
$\bar\nu_e \to \bar\nu_e$ oscillations at the solar $\delta m^2$
scale, the antineutrino mass-squared differences were adjusted to
account for the LSND and KamLAND oscillations (but not oscillations of
atmospheric antineutrinos)~\cite{cptv2}. An analysis of the
atmospheric data showed that the modified $CPT$-violating model did
not give a good fit and is excluded at the $3\sigma$
level~\cite{not-cptv}. Thus, neither four-neutrino models with $CPT$
conservation nor three-neutrino models with $CPT$ violation provides a
consistent explanation of all the data including LSND.

\section{Four neutrinos with $CPT$ violation}

The neutrino
flavor states $\nu_\alpha$ ($\alpha = e, \mu, \tau, s$) are related to
the mass eigenstates $\nu_i$ by a unitary matrix $U$, with
\begin{equation}
\nu_\alpha = \sum_{i=1}^4 U^*_{\alpha i} \nu_i \,. 
\label{eq:U}
\end{equation}
If $CPT$ is not conserved, then the corresponding unitary matrix $\bar
U$ for antineutrinos, given by
\begin{equation}
\bar\nu_\alpha = \sum_i \bar U_{\alpha i} \bar\nu_i \,,
\label{eq:Ubar}
\end{equation}
will not necessarily be equal to $U$. Furthermore, the neutrino and
antineutrino eigenmasses will not necessarily be the same. We will
assume that there are three different mass-squared difference scales
for neutrinos, $\delta m^2_s \ll \delta m^2_a \ll \delta m^2_L$, that
can explain the solar, atmospheric and LSND data, respectively, and
that the corresponding mass-squared differences for antineutrinos are
similar to those for neutrinos, {\it i.e.}, $\delta \bar m^2_s \approx
\delta m^2_s$, $\delta \bar m^2_a \approx \delta m^2_a$, and $\delta
\bar m^2_L \approx \delta m^2_L$.

\subsection{3~+~1 models}

In 3~+~1 models there is one neutrino mass well-separated from the
others by $\delta m^2_L$, and the sterile neutrino couples strongly
only to the isolated state. There are four different mass spectra in
3~+~1 models, depending on whether the isolated state is above or
below the others, and whether the other three neutrino states have a
normal or inverted mass hierarchy. We consider the case with $m_4 >
m_1, m_2, m_3$ and normal hierarchy, which implies $\delta m^2_{41}
\simeq \delta m^2_{42} \simeq \delta m^2_{43} = \delta m^2_L \gg
\delta m^2_{31} \simeq \delta m^2_{32} = \delta m^2_a \gg \delta
m^2_{21} = \delta m^2_s$, with similar relations for $\delta \bar
m^2_{ij}$ (the argument is similar for the other cases). Then the
relevant oscillation probabilities for neutrinos are approximately,
\begin{equation}
P(\nu_e \to \nu_e)_{\rm solar} \simeq
1 - 4 |U_{e1}|^2 |U_{e2}|^2 \sin^2\Delta_s \,, 
\label{eq:Pees}
\end{equation}
\begin{equation}
P(\nu_\mu \to \nu_\mu)_{\rm atm} \simeq
1 - 4 |U_{\mu3}|^2 (1 - |U_{\mu3}|^2 - |U_{\mu4}|^2)\sin^2\Delta_a \,,
\label{eq:Pmma} 
\end{equation}
\begin{equation}
P(\nu_\mu \to \nu_e)_{\rm MiniBooNE} \simeq
4 |U_{e4}|^2 |U_{\mu 4}|^2 \sin^2\Delta_L \,,
\label{eq:PmeL}
\end{equation}
\begin{equation}
P(\nu_\mu \to \nu_\mu)_{\rm CDHSW} \simeq
1 - 4 |U_{\mu4}|^2 (1 - |U_{\mu4}|^2) \sin^2\Delta_L \,,
\label{eq:PmmL}
\end{equation}
where $\Delta_x = \delta m^2_x L/(4E_\nu)$ is the usual oscillation
argument for $x = s$, $a$, or $L$. The relevant oscillation
probabilities for antineutrinos are approximately,
\begin{equation}
\bar P(\bar\nu_e \to \bar\nu_e)_{\rm KamLAND} \simeq
1 - 4 |\bar U_{e1}|^2 |\bar U_{e2}|^2 \sin^2\bar\Delta_s \,,
\label{eq:Pbareek}
\end{equation}
\begin{equation}
\bar P(\bar\nu_\mu \to \bar\nu_\mu)_{\rm atm} \simeq
1 - 4 |\bar U_{\mu3}|^2 (1 - |\bar U_{\mu3}|^2 - |\bar U_{\mu4}|^2)
\sin^2\bar\Delta_a \,,
\label{eq:Pbarmma}
\end{equation}
\begin{equation}
\bar P(\bar\nu_\mu \to \bar\nu_e)_{\rm LSND} \simeq
4 |\bar U_{e4}|^2 |\bar U_{\mu4}|^2 \sin^2\bar\Delta_L \,,
\label{eq:PbarmeL}
\end{equation}
\begin{equation}
\bar P(\bar\nu_e \to \bar\nu_e)_{\rm Bugey} \simeq
1 - 4 |\bar U_{e4}|^2 (1 - |\bar U_{e4}|^2) \sin^2\bar \Delta_L \,,
\label{eq:PbareeL}
\end{equation}
where $\bar\Delta_x = \delta \bar m^2_x L/(4E_\nu)$.  We note that
LSND has already made a measurement for $P_{\rm miniBooNE}$, although
at the $2\sigma$ level it is consistent with both $\bar P_{\rm LSND}$
and zero.

If $CPT$ is not conserved\footnote{The reason that 3~+~1 models 
with $CPT$ conservation
(where $\bar U = U$) are disfavored
is that $|U_{\mu 4}|^2$ must be small from a
combination of the strict CDHSW limit~\cite{cdhsw} and the large
mixing of atmospheric $\nu_\mu$, and $|U_{e4}|^2$ must be small from a
combination of the strict Bugey limit~\cite{bugey} and the large
mixing of solar $\nu_e$ and KamLAND $\bar\nu_e$, leading to an
upper bound on the LSND amplitude $4 |U_{e4}|^2
|U_{\mu4}|^2$~\cite{4nu-constraints, our-four}. The CDHSW and Bugey
limits are $\delta m^2_L$ dependent (as is the LSND allowed
amplitude), but a comparison at all $\delta m^2_L$ shows that nowhere
does the LSND+KARMEN 95\%~C.~L. allowed region overlap 
the 95\%~C.~L. allowed
region from the other experiments~\cite{grimus-schwetz}. 
The most recent
comprehensive analysis concludes that the 3~+~1 models have a goodness
of fit of at most $5.6 \times 10^{-3}$~\cite{not-four}. }, 
then in general $\bar U \ne U$. The
Bugey+CDHSW bound on LSND and MiniBooNE can now be evaded since Bugey
limits $|\bar U_{e4}|$ and CDHSW limits $|U_{\mu4}|$, but $|\bar
U_{\mu4}|$ and $|U_{e4}|$ are no longer tightly constrained. The
bounds on the amplitudes for $\bar\nu_\mu \to \bar\nu_e$ and $\nu_\mu
\to \nu_e$ oscillations must be determined 
separately\footnote{Although neutrino telescopes are 
capable of probing the LSND scale~\cite{renata}, 
they are unable to test $CPT$-violating schemes
because neutrino and antineutrino oscillation 
probabilities are not measured separately.}:

\begin{enumerate}

\item[(i)] The best constraint on $|U_{e4}|$ comes from $\nu_e$
disappearance performed during GALLEX testing with a $^{51}$Cr
neutrino source~\cite{hampel}; it does not extend as low in $\delta
m^2_L$ as the Bugey constraint and is about an order of magnitude
weaker at high $\delta m^2_L$. 
(The GALLEX survival probability is given by Eq.~\ref{eq:PbareeL} 
with $\bar U_{e4}$ replaced 
by $U_{e4}$). When combined with the limit on
$|U_{\mu4}|$ from CDHSW~\cite{cdhsw}, a range of $\nu_\mu \to
\nu_e$ oscillation amplitudes ($4 |U_{e4}|^2 |U_{\mu4}|^2$) 
is excluded; 
see Fig.~\ref{fig:3+1nu}. A positive MiniBooNE
result that exceeds the upper bound in Fig.~\ref{fig:3+1nu} will rule
out 3~+~1 models with $CPT$ violation, while a null or positive 
 result that obeys the upper bound is easily accommodated. 
A MiniBooNE measurement consistent with
the region allowed by LSND and KARMEN would mean that the values of
$4 |U_{e4}|^2 |U_{\mu4}|^2$ and $4 |\bar U_{e4}|^2 |\bar U_{\mu 4}|^2$
are similar, but the Bugey+CDHSW bound demands $\bar U \neq U$. 
In fact, a MiniBooNE measurement that lies below the GALLEX+CDHSW
upper limit
and above the Bugey+CDHSW limit, would place a lower bound on 
the amount of $CPT$ violation.

\begin{figure}[t]
\centering\leavevmode
\includegraphics[width=3.2in]{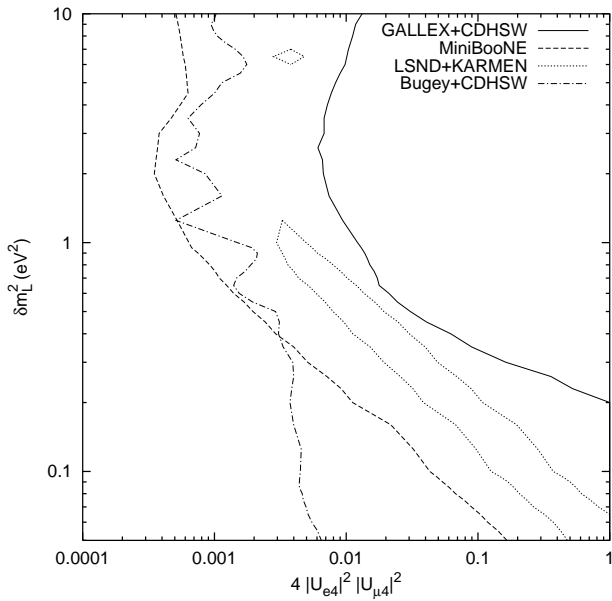}
\caption[]{Upper bound (solid) on the $\nu_\mu \to \nu_e$ oscillation
amplitude $4 |U_{e4}|^2 |U_{\mu4}|^2$ from the GALLEX limit on 
$|U_{e4}|$ and the CDHSW 
limit on $|U_{\mu4}|$ (90\%~C.~L.
results are used in both cases). The dot-dashed line is the 99\% C.~L. 
upper bound from Bugey and CDHSW if $CPT$ is conserved~\cite{grimus-schwetz}.
Also shown are the expected sensitivity (dashed) of
the MiniBooNE experiment  and, for comparison, the
allowed region (within the dotted lines) 
for $4 |\bar U_{e4}|^2 |\bar U_{\mu4}|^2$ from a
combined analysis of LSND and KARMEN data, both at 
the 90\% C.~L~\cite{church}. 
\label{fig:3+1nu}}
\end{figure}

\item[(ii)] The best accelerator constraint on $|\bar U_{\mu4}|$ comes
from the CCFR experiment that searched for $\bar\nu_\mu$
disappearance~\cite{ccfr2}; there is no limit for $\delta m^2_L <
7$~eV$^2$. Although $|\bar U_{\mu4}|$ cannot be so large as to disturb
the usual fits to atmospheric data (which indicate the dominant
oscillation is at the $\delta m^2_a$ scale), $|\bar U_{\mu4}|^2$ can
probably be of order a few per cent, which allows the LSND amplitude
$4 |\bar U_{e4}|^2 |\bar U_{\mu4}|^2$ to be large enough to account
for the LSND $\bar\nu_\mu \to \bar\nu_e$ data, at least for the larger
values of $\delta \bar m^2_L$ allowed by LSND 
(near 1~eV$^2$ and 7~eV$^2$);
see Fig.~\ref{fig:3+1nubar}.

\begin{figure}[t]
\centering\leavevmode
\includegraphics[width=3.2in]{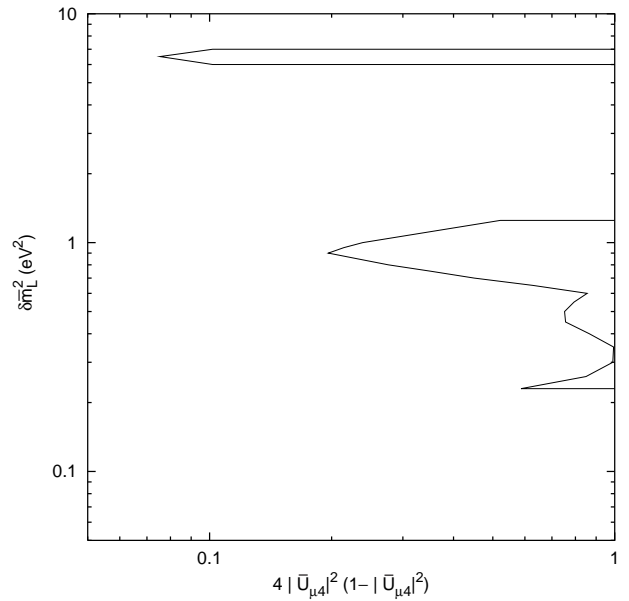}
\caption[]{Lower bounds on 
$4 |\bar U_{\mu4}|^2 (1 - |\bar U_{\mu4}|^2)$
(the amplitude for atmospheric $\bar\nu_\mu$ survival at the LSND mass
scale) from the Bugey limit on $\bar\nu_e$ disappearance and the
$\bar\nu_\mu \to \bar\nu_e$ oscillation amplitude indicated by LSND and
KARMEN (90\%~C.~L. results are used in both cases).
\label{fig:3+1nubar}}
\end{figure}

\end{enumerate}

We note that fits for neutrinos and antineutrinos no longer must
agree, so that, {\it e.g.}, the solar fits are now independent of the
KamLAND fits. Thus, any differences that might occur between the
KamLAND and solar neutrino allowed regions could be explained by $CPT$
violation. Similarly, results for atmospheric neutrinos and
antineutrinos (which could perhaps be measured separately in
MINOS~\cite{minos}) could also be different and still be easily
accommodated in the model. On the other hand, if there were no
discernible difference between the neutrino and antineutrino fits to
solar, KamLAND, and atmospheric data, the amount of $CPT$ violation
needed would be small, since then the only differences between 
neutrino and antineutrino parameters would occur in the small mixings
between the sterile and the active states.

\subsection{2~+~2 models}

Constraints on 2~+~2 models are different from those on the 3~+~1
models because they have different
expressions for the oscillation probabilities. The
difficulty with these models is that the sterile neutrino is strongly
coupled to solar and/or atmospheric neutrino oscillations, and there
are bounds on the amount of sterile content in each case. In 2~+~2
models, solar $\nu_e$ oscillate predominantly to a linear combination
of $\nu_\tau$ and $\nu_s$ and atmospheric $\nu_\mu$ oscillate to the
orthogonal combination~\cite{our-four}:
\begin{eqnarray}
\nu_e &\to& -\sin\alpha\,\nu_\tau + \cos\alpha\,\nu_s \,,
\label{eq:nue}\\
\nu_\mu &\to& \cos\alpha\,\nu_\tau + \sin\alpha\,\nu_s \,.
\label{eq:numu}
\end{eqnarray}
If $CPT$ is violated, then there is a similar sterile mixing angle
$\bar\alpha$ for antineutrinos. The amount of sterile content is
$\cos^2\alpha$ in solar neutrino oscillations, $\cos^2\bar\alpha$ in
KamLAND, $\sin^2\alpha$ in atmospheric neutrino oscillations, and
$\sin^2\bar\alpha$ in atmospheric antineutrino oscillations. Fits to
solar neutrino data give the 99\%~C.~L. limit~\cite{not-four},
\begin{equation}
\cos^2\alpha \le 0.45 \,;
\label{eq:solarlimit}
\end{equation}
note that there is no limit on $\cos^2\bar\alpha$ from KamLAND
since the short baseline has negligible matter effects and it
therefore does not test the the sterile content. Fits to the
atmospheric neutrino data (which do not distinguish between neutrinos
and antineutrinos) give the 99\%~C.~L. limit~\cite{not-four},
\begin{equation}
{2\over3}\sin^2\alpha + {1\over3}\sin^2\bar\alpha \le 0.35 \,,
\label{eq:atmoslimit}
\end{equation}
due to the lack of matter effects that would occur in $\nu_\mu \to
\nu_s$ oscillations~\cite{superK-tau}. In Eq.~(\ref{eq:atmoslimit}),
neutrinos contribute with twice the strength of antineutrinos
because of their larger interaction cross section.

\begin{figure}[t]
\centering\leavevmode
\includegraphics[width=3.2in]{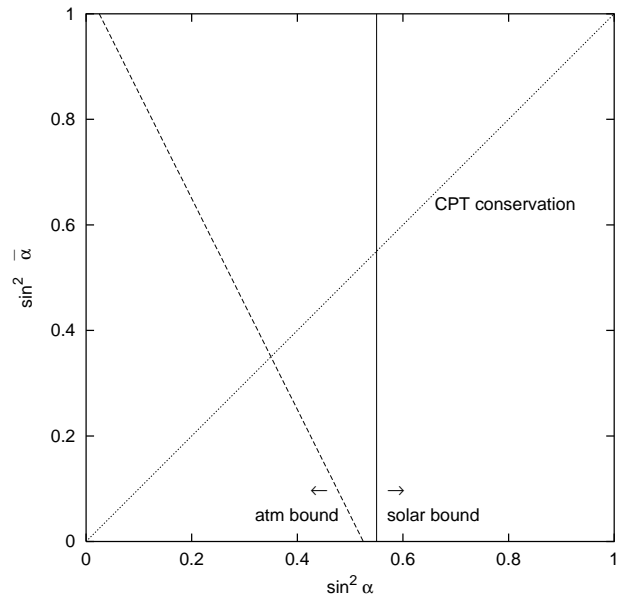}
\caption[]{Constraints on sterile neutrino mixing angles $\alpha$ and
$\bar\alpha$ from solar (solid) and atmospheric (dashed) data.
The dotted line is the prediction if $CPT$ is conserved.
\label{fig:2+2bounds}}
\end{figure}

The bounds on 2~+~2 models are shown in Fig.~\ref{fig:2+2bounds}
versus $\sin^2\alpha$ and $\sin^2\bar\alpha$. The $CPT$-conserving
case $\alpha = \bar\alpha$ is indicated by the dotted line. A recent
comprehensive analysis of 2~+~2 models with $CPT$ conservation
concludes that the goodness of fit to all data is only $1.6 \times
10^{-6}$~\cite{not-four}, which is worse than that of 3~+~1 models
with $CPT$ conservation\footnote{The strong exclusion of the $CPT$
conserving case is also evident by adding together the constraints of
Eqs.~(\ref{eq:solarlimit}) and (\ref{eq:atmoslimit}), which for
$\bar\alpha = \alpha$ yields $\cos^2\alpha + \sin^2\alpha \le
0.80$.}. When $CPT$ is violated, {\it i.e.}, $\alpha \ne \bar\alpha$, there
is no region that obeys both the solar and atmospheric bounds. Thus even
when $CPT$ is violated, 2~+~2 models are strongly disfavored.

The analysis that lead to Eq.~(\ref{eq:solarlimit}) used the standard
solar model (SSM)~\cite{bahcall} 
neutrino fluxes, including their theoretical
uncertainties. If the $^8$B neutrino flux is allowed to be free, 
this bound relaxes to $\cos^2\alpha \le 0.61$~\cite{not-four2}. Then
the solar constraint in Fig.~\ref{fig:2+2bounds} is $\sin^2\alpha \ge
0.39$, and there is a small allowed region with $0.39 \le \sin^2\alpha
\le 0.53$ and small $\sin^2\bar\alpha$. This can be understood
qualitatively as follows: although there is no room for a full
sterile neutrino in the solar and atmospheric data when $CPT$ is
conserved, with $CPT$ violation the fact that KamLAND does not test
the sterile content means that oscillations of atmospheric
antineutrinos can be largely to active neutrinos, which effectively
dilutes the constraint on the sterile content in atmospheric neutrino
oscillations. However, there is currently no reason to believe that
the $^8$B neutrino flux is not well-described by the SSM.

\subsection{Hybrid models}

Since $CPT$ is violated, in principle one can have a 3~+~1 model in
the neutrino sector and a 2~+~2 model in the antineutrino sector, or
vice versa. We now examine these two possibilities:

\begin{enumerate}

\item[(i)] (3~+~1)$_\nu$, (2~+~2)$_{\bar\nu}$. In this case,
oscillations in the neutrino sector are predominantly to active
neutrinos, and there is no sterile content bound from solar neutrino
data. For the atmospheric data, this situation is equivalent to
$\alpha \simeq 0$ in Eq.~(\ref{eq:atmoslimit}), which leads 
to no bound on $\bar\alpha$. 
Observation of a non-negligible sterile content in the oscillations of 
atmospheric antineutrinos, but not atmospheric neutrinos, 
would provide supporting evidence for this model, 
although such sterile content is not required.

\item[(ii)] (2~+~2)$_\nu$, (3~+~1)$_{\bar\nu}$. In this case,
oscillations in the antineutrino sector are predominantly to active
species. The bound in Eq.~(\ref{eq:solarlimit}) remains the same
($\sin^2\alpha \ge 0.55$), and the bound from the atmospheric data can
be found by setting $\bar\alpha = 0$, {\it i.e.}, 
$\sin^2\alpha \le 0.53$. Therefore this hybrid combination, 
like the pure 2~+~2 model, is strongly disfavored.

\end{enumerate}

Furthermore, it is possible that more general models that do not fit
the 3~+~1 and/or 2~+~2 structure can provide a satisfactory fit to the
data, although we do not perform such an analysis here.

\section{Summary}

We have argued that a four-neutrino model with $CPT$ violation can
provide an explanation of all neutrino oscillation data 
if the neutrino sector has a 3~+~1 structure and the 
antineutrino sector is either 3~+~1 or 2~+~2; a 2~+~2 structure is
not allowed in the neutrino sector. 
 If the antineutrino sector is 3~+~1, then
$\delta \bar m^2_L$ must be near 1 or 7~eV$^2$, while if it is 2~+~2, 
the entire range of $\delta \bar m^2_L$ allowed by LSND, KARMEN and 
Bugey is possible. 
 
A detection of $\nu_\mu \to \nu_e$
oscillations by MiniBooNE with parameters 
 above the solid line in Fig.~\ref{fig:3+1nu} will require 
$CPT$ violation to be too large and 3~+~1 models 
with $CPT$ violation will be excluded.
However, if MiniBooNE has a null oscillation result 
or finds oscillation parameters that lie below the solid line,
such models will remain viable.
Even if the MiniBooNE allowed region 
for neutrinos falls within the LSND+KARMEN 
allowed region for antineutrinos, 
the Bugey+CDHSW bound on $CPT$ conserving 3~+~1 models (dot-dashed line
in  Fig.~\ref{fig:3+1nu}) implies 
 $CPT$ violation of the size
of the LSND effect (${\cal{O}}(10^{-3})$ in terms of 
oscillation probabilities). 

\vskip 0.1in
{\it Acknowledgments.}
 This research was supported
in part by the U.S. Department of Energy under Grants
No.~DE-FG02-95ER40896, DE-FG02-01ER41155 and DE-FG02-91ER40676, 
and in part by the Wisconsin Alumni Research Foundation.

%%REFERENCES

\end{document}